\documentclass[letterpaper,journal]{IEEEtran}
\usepackage{amsmath,amsfonts}
\usepackage{array}
\usepackage[shrink=10,stretch=10]{microtype}
\usepackage{textcomp}
\usepackage{stfloats}
\usepackage{url}
\usepackage{verbatim}
\usepackage{graphicx}
\usepackage[hidelinks]{hyperref}
\usepackage{cite}
\usepackage[british]{babel}

\newtheorem{defi}{Definition}
\newtheorem{prop}{Proposition}
\newtheorem{ex}{Example}

\newcommand{\orcidID}[1]{
    \unskip\hspace{0.1em}\href{https://orcid.org/#1}{\raisebox{-0.12\height}{\includegraphics[height=0.8em]{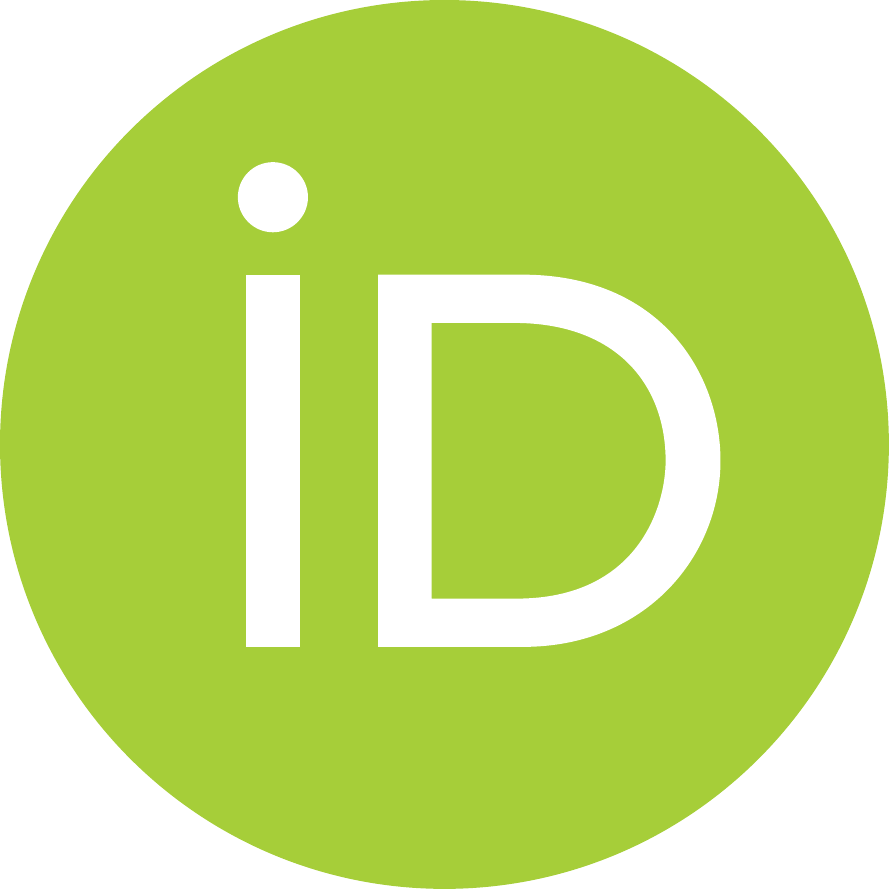}}}%
}

\begin{document}

\title{Private Randomness Agreement and its Application in Quantum Key Distribution Networks}

\author{René Bødker Christensen\orcidID{0000-0002-9209-3739} and Petar Popovski\orcidID{0000-0001-6195-4797}%
    \thanks{This paper was supported in part by the Villum Investigator Grant ``WATER'' from the Velux Foundations, Denmark.}%
    \thanks{R.B. Christensen is with the Department of Mathematical Sciences, Aalborg University, and Department of Electronic Systems, Aalborg University. (e-mail: \texttt{rene@math.aau.dk})}%
    \thanks{P. Popovski is with the Department of Electronic Systems, Aalborg University. (e-mail: \texttt{petarp@es.aau.dk})}}

\markboth{Article Postprint}%
{Christensen \MakeLowercase{\textit{et al.}}: Private Randomness Agreement}

\IEEEpubid{%
    \begin{minipage}{\textwidth}
      \vglue 1cm
      \centering
      ©2022 IEEE. Personal use of this material is permitted. Permission from IEEE must be obtained for all other uses, in any current or future media, including reprinting/republishing this material for advertising or promotional purposes, creating new collective works, for resale or redistribution to servers or lists, or reuse of any copyrighted component of this work in other works. DOI: \href{https://doi.org/10.1109/LCOMM.2022.3225262}{10.1109/LCOMM.2022.3225262}
    \end{minipage}%
}

\maketitle

\begin{abstract}
  We define a variation on the well-known problem of private message transmission. This new problem called \emph{private randomness agreement} (PRA) gives two participants access to a public, authenticated channel alongside the main channels, and the `message' is not fixed a priori.
  Instead, the participants aim to agree on a random string completely unknown to a computationally unbounded adversary.
  We define privacy and reliability, and show that PRA cannot be solved in a single round. We then show that it can be solved in three rounds, albeit with exponential cost, and give an efficient four-round protocol based on polynomial evaluation.
\end{abstract}


\section{Introduction}
Exchanging key material is a fundamental problem in cryptography.
By using quantum mechanics, the so-called BB84 protocol \cite{bb84} allows two parties connected via a direct quantum link to establish shared material for generating a key in such a way that eavesdroppers are detected with overwhelming probability.
Over large distances, however, the achievable key rate of direct, repeaterless quantum links drops dramatically\footnote{E.g. requiring almost $7\cdot 10^9$ channel uses per secret bit if a 500km optical link with a loss of 0.2dB/km is used.}~\cite{Pirandola2017}.

When no direct quantum link exists between parties it has been suggested -- see e.g. \cite[Sec. 4.2]{MehicEtAl} and \cite{GDKH21} -- that they instead route their key material through a quantum key distribution (QKD) network, in which each connection is a direct quantum link. Thereby, the parties can exchange keys securely as long as each node in the QKD network is trusted.
If any node in the QKD network is malicious, the security breaks down as this node learns any key material routed through it. Furthermore, it could potentially alter the key, so the parties exchanging keys end up with different keys.

Taking this problem as our point of departure, we define a more general problem that can be seen as a variation on the problem of \emph{private message transmission} (PMT), which has been extensively studied~\cite{Spini,Patra,KurosawaSuzuki,KurosawaSuzuki09,WangDesmedt,WangDesmedt01,DolevDwork}. In PMT, two participants, Alice and Bob, are connected via $n$ channels, some of which are controlled by an adversary. The exact channels under adversarial control, referred to as the \emph{corrupted} channels, are unknown to Alice and Bob, however. The aim is to devise some strategy on how to use these channels to allow Alice to send a message to Bob privately and reliably -- even if the adversary alters the information sent across the corrupted channels.

Returning to the QKD-setting, we can solve the QKD problem using PMT. Namely, if the QKD-network contains $n$ vertex-disjoint paths between Alice and Bob, we can consider these as the $n$ channels in the PMT-setting. The secret message is then the secret key chosen by Alice.
Thus, the existing literature on PMT~\cite{Spini,Patra,KurosawaSuzuki,KurosawaSuzuki09,WangDesmedt,WangDesmedt01,DolevDwork} provides a solution to the QKD-problem as long as the power of the adversary is sufficiently limited.
We argue, however, that the model used in PMT is overly restrictive for the QKD-setting. The reason for this is twofold. First, in PMT the two parties cannot communicate outside the $n$ channels, whereas the goal in the QKD setting is to exchange a key which can later be used to communicate securely via \emph{another} channel, which is public and authenticated. Otherwise, they might as well use the QKD-network to transmit the message itself rather than a key to encrypt the message later. Second, in PMT Alice has a specific message that she wants to transmit to Bob. In QKD, Alice and Bob still need to exchange information to establish a shared key, but not \emph{one specific} key -- they need only agree on \emph{some} random key.
As we will show in this work, this changes the feasibility and optimality of the problem.

The public channel introduced in this work illustrates an interesting point in relation to QKD: the secrecy in QKD relies on the physical properties of the quantum channel such as non-cloneability, superpositions, and entanglement. As we demonstrate in this work, a classical public channel can amplify the security of the quantum channel, improving the overall secrecy of the system.

\IEEEpubidadjcol
\section{Problem definition}\label{sec:problemDef}
We now define the details of our model. The two participants, Alice and Bob, are connected via $n$ channels, which will be indexed by $1,2,\ldots,n$. We will refer to these as the \emph{main channels}. In addition, they are connected via a public, authenticated channel which we will denote by index $0$. A computationally unbounded and active adversary $\mathcal{A}$ controls $t<n$ of the main channels. That is, $\mathcal{A}$ sees the information sent across the channels under its control, and $\mathcal{A}$ being active means that it can also alter this information or block it completely.\footnote{Otherwise, $\mathcal{A}$ would be \emph{passive}. This has a trivial solution: Alice samples a random key $k\in\mathbb{F}_q^\ell$, creates an additive sharing $k=\sum_{i=1}^n k_i$, and then routes $k_i$ via the $i$'th channel. When $\mathcal{A}$ is passive, Bob is guaranteed to receive the original shares sent by Alice, meaning that he can always reconstruct $k$. The key $k$ remains private by the $(n-1)$-privacy of additive secret sharing.}
Furthermore, the adversary also receives the information sent across the public channel, but cannot alter or block this in any case.\footnote{If the adversary could tamper with the public channel, it would simply be another channel with the same properties as the main channels. That is, the problem reduces to standard PMT.}
The goal for Alice and Bob is to transmit data via the channels in a number of rounds such that they at the end know a shared random key, that is completely unknown to the adversary. As is done in~\cite{DolevDwork}, each round allows Alice to send information to Bob, or Bob to send information to Alice, but not both within the same round.
We will refer to this problem as \emph{private randomness agreement} (PRA). Note that the BB84 protocol~\cite{bb84} also aims to provide private randomness, but in the setting where Alice and Bob are connected via a direct quantum link.

Regarding the number of corrupted channels, we consider the case $t=n-1$, which is the worst case.\footnote{Clearly, the case $t=n$ is impossible to solve since an adversary could block all channels under its control.} As we will show, the problem does actually have a solution in this case. This highlights a key difference between PMT and PRA, as it is well-known that PMT cannot be solved for $2t\geq n$~\cite[Thm.\,5.2]{DolevDwork} in the case of actively corrupted channels. Thus, the changes introduced in PRA allows us to tolerate a much more powerful adversary.

We now turn to defining precisely what we mean by privacy and reliability in the PRA-setting. The definitions resemble those used in PMT\cite{DolevDwork,Franklin2000}, but there is an important difference in the case of privacy. In PMT, a protocol being private essentially means that the adversary cannot distinguish between protocol transcripts when sending $m$ and when sending $m'$. But in that setting, the intended message $m$ is known \emph{a priori}. In PRA, Alice and Bob have not decided on a desired key beforehand. Furthermore, if the protocol fails, and Alice and Bob output different keys, it is unclear what should be considered the `intended' key in that execution. Thus, we will consider transcripts conditioned on the output of both Alice and Bob.
More precisely, let $\Pi$ be a PRA-protocol, and assume that Alice and Bob will output keys $k_A$ and $k_B$, respectively, where $k_A,k_B$ are both sampled from a set $\mathcal{K}$ of possible keys.
Let $\mathcal{A}$ be an adversary and assume that its input randomness is $r$. That is, the random choices made by $\mathcal{A}$ is specified by $r$ similarly to how the random tape is used to define probabilistic Turing machines.
For such an adversary and a protocol $\Pi$, we then define the random variable $\mathcal{A}_\Pi(k_A,k_B,r)$ describing the transcript of $\Pi$ from the view of $\mathcal{A}$ conditioned on the output of Alice being $k_A$ and the output of Bob being $k_B$. The probability distribution of this variable depends on the randomness of both Alice and Bob.

In the following definition and throughout the work, we assume that the key space $\mathcal{K}$ has the structure of an additive group. This is not an unreasonable assumption if the key is later to be used as a one-time pad. With this assumption, addition of keys in $\mathcal{K}$ is well-defined and gives another key in $\mathcal{K}$.
\begin{defi}\label{defi:protocolSecurity}
  A PRA-protocol $\Pi$ is called $\delta$-reliable if the following holds.
  \begin{itemize}
    \item $\Pi$ is perfectly private. That is, for any $k_A,k_A',e\in\mathcal{K}$ the transcripts $\mathcal{A}_\Pi(k_A,k_A+e,r)$ and $\mathcal{A}_\Pi(k_A',k_A'+e,r)$ are perfectly indistinguishable.
    \item When using $\Pi$, the probability that Alice and Bob output different keys is at most $\delta$.
  \end{itemize}
\end{defi}
In essence, the guarantee provided by Definition~\ref{defi:protocolSecurity} is that the adversary cannot learn the specific keys that Alice and Bob output. It may learn whether the protocol has succeeded or not as well as the difference $e=k_B-k_A$, but the reliability requirement ensures that $e=0$ except with some (typically very small) probability.
Note that this is in line with the security in PMT: we require Alice's message to remain private, but do not guarantee anything about the output of Bob when reliability fails.

We note here that it is possible to define a stronger privacy notion, namely to require $\mathcal{A}_\Pi(k_A,k_B,r)\sim\mathcal{A}_\Pi(k_A',k_B',r)$ for any $k_A,k_A',k_B,k_B'\in\mathcal{K}$. This stronger privacy notion implies that the adversary is not even allowed to learn whether the protocol has succeeded or not -- and hence it also learns  nothing about the difference $k_B-k_A$. This seems to be a much more difficult problem to solve, so we limit ourselves to the security provided by Definition~\ref{defi:protocolSecurity}.

We note that the statement of protocol privacy is somewhat verbose. In particular, perfect privacy implies that for any error $e$, the transcript of the adversary must be distributed in the same way regardless of the output of Alice, $k_A$. Thus, we can drop $k_A$ from the notation, and simply use $\mathcal{A}_\Pi(e,r)$ to denote the random transcript in an execution of $\Pi$.
Translating these observations to statements on entropy, the protocol $\Pi$ being perfectly private implies
\begin{equation}\label{eq:entropy}
  \begin{aligned}
  H(K_A\mid \mathcal{A}_\Pi(e,r))&=H(K_A)\\
  H(K_B\mid \mathcal{A}_\Pi(e,r))&=H(K_B),
  \end{aligned}
\end{equation}
where $H$ denotes the usual binary Shannon entropy~\cite{Shannon}, and $K_A$ and $K_B$ denote the random variables describing the output of Alice and Bob, respectively.

To illustrate the requirements of Definition~\ref{defi:protocolSecurity}, we consider a minimal example similar to the case of a passive adversary described above. This gives a protocol that is private against an active adversary, but provides virtually no reliability.
\begin{ex}
  Let $n=2$, meaning that the channels are the public channel, one honest channel, and one corrupt channel. Assume for clarity that the corrupt channel has index $2$. Consider a protocol where Alice samples a key $k_A\in\mathbb{F}_q^\ell$ uniformly at random, and creates a sharing $k_0+k_1+k_2=k_A$ of the key. She then sends $k_0$ across the public channel, and $k_i$ across the $i$'th main channel. Bob receives $k_0$, $\tilde{k}_1=k_1$, and $\tilde{k}_2$ where the latter may be altered by the adversary (but Bob does not know if $\tilde{k}_1$ or $\tilde{k}_2$ is the one provided by the adversary). Our toy-protocol specifies that he should then output $k_B=k_0+\tilde{k}_1+\tilde{k}_2$.

  Note now that regardless of the outputs of Alice and Bob, $k_A$ and $k_B$, the transcript $\mathcal{A}_\Pi(k_A,k_B,r)$ always comprises only $k_0$, $k_2$, and $\tilde{k}_2$. This specifies the difference $e=k_B-k_A=\tilde{k}_2-k_2$ chosen by the adversary.
  Further, $k_1$ acts as a one-time pad in the sharing of $k_A$, so the adversary learns nothing about $k_A$. In other words, it is impossible to distinguish $\mathcal{A}_\Pi(k_A,k_A+e,r)$ from $\mathcal{A}_\Pi(k_A',k_A'+e,r)$ for any choice of $k_A'\in\mathbb{F}_q^\ell$.

  It should be clear, however, that this protocol provides no reliability, as the adversary controls the difference between the output keys $k_A$ and $k_B$. Thus, this protocol is $1$-reliable.
\end{ex}

\section{Bounding the number of rounds}
We now consider the minimum number of rounds necessary to successfully establish a key. First, we rule out the possibility of one-round protocols. This shows another important difference between PMT and PRA, as PMT can be solved in one round as long as $2t<n$ and $\delta>0$ \cite[Thm.\,3.2]{WangDesmedt}, or if $3t<n$ and $\delta=0$ \cite[Thm.\,5.1]{DolevDwork}.

In this section, we will always assume that Alice sends information in the first round. Note that in the case of a two-round protocol, Bob will be sending information in the second round -- otherwise, we could consider it as a single round and reuse the arguments in Proposition~\ref{prop:oneRoundImpossible}.
\begin{prop}\label{prop:oneRoundImpossible}
  If the adversary controls $t=n-1$ out of $n$ channels, any $\delta$-reliable one-round PRA protocol has $\delta\geq 1-2^{-H(K_A)}$, where $K_A$ is the output of Alice.
\end{prop}
\begin{IEEEproof}
  Assume for contradiction that $\Pi$ is such a protocol, and assume without loss of generality that Alice transmits to Bob. Since $\Pi$ is one-round, the input randomness of Alice determines the key $K_A$ that Alice will output.

  For each $i\in\{1,2,\ldots,n\}$, let $X_i\in\mathcal{X}$ denote the random variable describing the information Alice transmits via the $i$'th channel in the network, and define $X_0$ to be the information transmitted via the public channel. By perfect privacy, we have
  \begin{equation}\label{eq:privacyBound}
    H(K_A\mid \{X_i\}_{i\neq j})=H(K_A)
  \end{equation}
  for any $j\in\{1,2,\ldots,n\}$. Now, consider an adversary that discards the $X_i$ sent by Alice and replaces it with a freshly sampled $\tilde{X}_i\leftarrow\mathcal{X}$ for each corrupted channel $i$. Since these $\tilde{X}_i$ are independent from all other variables, standard properties of conditional entropy~\cite{CoverThomas} imply that for each $i^\ast,j\in\{1,2,\ldots,n\}$ with $j\neq i^\ast$, we have
  \begin{align}\label{eq:learnNothing}
    H(K_A\mid X_0,X_{i^\ast},\{\tilde{X}_{i}\}_{i\neq i^\ast}) &=H(K_A\mid X_0,X_{i^\ast})\nonumber\\
                                                               &\geq H(K_A\mid X_0,\{X_i\}_{i\neq j})\nonumber\\
                                                               &=H(K_A),
  \end{align}
  where the last equality follows from~\eqref{eq:privacyBound}.
  But letting $i^\ast$ denote the honest channel, Bob learns exactly the values of $X_0,X_{i^\ast},\{\tilde{X}_i\}_{i\neq i^\ast}$, which by~\eqref{eq:learnNothing} provides no information on the key $K_A$, which Alice outputs.
  This proves the theorem.
\end{IEEEproof}
Despite our best efforts, we have not been able to disprove the possibility of two-round protocols. We can, however, simplify the statement that needs to be proved. Namely, it is only necessary to rule it out for $n=2$ channels, as the following theorem ensures that this rules out two-round protocols in general.
\begin{prop}
  If PRA can be solved in two rounds for $n>2$ channels, then it can also be solved in two rounds for $n=2$ channels.
\end{prop}
\begin{IEEEproof}
  Let $\Pi$ be a two-round solving PRA for $n>2$ channels. Consider the protocol $\Pi'$ for two channels where Alice and Bob follow $\Pi$, but discard anything that is to be sent across channels $3,4,\ldots,n$ in $\Pi$.
  Alice and Bob's transcripts during $\Pi'$ are equivalent to their transcripts during $\Pi$ against an adversary that always blocks channels $3,4,\ldots,n$. The claim then follows from the security of $\Pi$.
\end{IEEEproof}
In order to give an upper bound on the minimal number of rounds, we present a protocol that is secure, but computationally inefficient. Building on a cut-and-choose argument, the protocol is conceptually simple: Alice and Bob exchange a large number of keys and then sacrifice all but one.
The precise description of the proposed method is given in Figure~\ref{prot:active} on page~\pageref{prot:active}.
\begin{figure*}
  \small
  \rule{\linewidth}{1.5pt}
  \begin{center}
    \unskip\textbf{Protocol~\ref{prot:active}}
  \end{center}
  \unskip
  \begin{enumerate}
    \item Alice samples $2^\sigma$ random strings $x_i\in\mathbb{F}_q^\ell$ independently and uniformly at random, and creates additive secret sharings $x_i=\sum_{j=1}^n x_{ij}$. She then routes $(x_{ij})_{i=1,2,\ldots,2^\sigma}$ via the $j$'th channel.
    \item\label{item:chooseIndex}
    Bob receives the possibly altered shares $\tilde{x}_{ij}$ and computes $\tilde{x}_i=\sum_{i=1}^n \tilde{x}_{ij}$. He chooses uniformly at random an index $s\in\{1,2,\ldots,2^\sigma\}$, and sends $s$ as well as $(\tilde{x}_{ij})_{i\neq s}$ across the public channel.
    \item Alice performs the check $\tilde{x}_{ij}=x_{ij}$ for every $i\neq s$ and every $j\in\{1,2,\ldots,n\}$. She defines the set
    \begin{equation*}
      \mathcal{T}=\{j\in\{1,2,\ldots,n\}\mid \forall i\neq s\colon \tilde{x}_{ij}=x_{ij}\},
    \end{equation*}
    and outputs $k_A=\sum_{j\in\mathcal{T}}x_{sj}$. Via the public authenticated channel she sends $\mathcal{T}$ to Bob, who outputs $k_B=\sum_{j\in\mathcal{T}}\tilde{x}_{sj}$.
  \end{enumerate}
  \rule{\linewidth}{1.5pt}
  \caption{Three-round key exchange against active and computationally unbounded adversaries.
      After performing this PRA protocol for $n$ channels, Alice and Bob will share a key unknown to the adversary. Before executing the protocol, they have agreed on a security parameter $\sigma$.}
  \label{prot:active}
\end{figure*}
\begin{prop}\label{prop:threeRoundSecure}
  Protocol~\ref{prot:active} is $\delta$-reliable for $\delta=(n-1)2^{-\sigma}$, where $\sigma$ is a security parameter.
\end{prop}
\begin{IEEEproof}
  In order for Alice and Bob to output different keys $k_A\neq k_B$, there must be some $j^\ast\in\mathcal{T}$ such that $\tilde{x}_{sj^\ast}\neq x_{sj^\ast}$. By definition of $j^\ast$, we have $\tilde{x}_{ij^\ast}=x_{ij^\ast}$ for every $i\neq s$.
  Thus, considering a corrupted channel $j$ in isolation, the best the adversary can do is to leave $x_{ij}$ unchanged for all but one index $i$. This is because this maximizes the probability that, after Step~\ref{item:chooseIndex} in Protocol~\ref{prot:active}, Alice will treat channel $j$ as honest, and Bob will include $j\in\mathcal{T}$ with probability $2^{-\sigma}$. By choosing different indices $i$ to alter for each corrupted channel $j$, the probability that the adversary succeeds in getting $j^\ast\in\mathcal{T}$ is at most $(n-1)2^{-\sigma}$ by the union bound. This proves the first claim.

  For the privacy part, denote the honest channel by $j^\ast$. Recall that we consider transcripts for a fixed input randomness of the adversary. Thus, observe that the difference $E=K_B-K_A$ in an execution of the protocol as well as $\mathcal{T}$ are completely determined by $S$, $(X_{ij})_{j\neq j^\ast}$, and $(\tilde{X}_{ij})_{j\neq j^\ast}$. In addition, once the adversary's randomness is fixed, the strings $(\tilde{X}_{ij})_{j\neq j^\ast}$ are completely determined by this randomness and $(X_{ij})_{j\neq j^\ast}$.
  Hence, if we fix a difference $E=e$, only a subset of the tuples $(S,(X_{ij})_{j\neq j^\ast})$ are possible. We note, however, that since both $S$ and the individual $X_{ij}$ are uniformly random, each tuple is equally likely to appear in a transcript $\mathcal{A}(k_A,k_A+e,r)$ for a given key $k_A$.
  
  Finally, the adversary never learns the value of $X_{sj^\ast}$ which is again uniformly random. This implies that the transcript $\mathcal{A}(k_A,k_A+e,r)$ mentioned above can correspond to any possible key $K_A$, and any such key is equally likely.
\end{IEEEproof}
We note that it is possible to reduce the communication cost in Step~\ref{item:chooseIndex} of Protocol~\ref{prot:active}. Namely, by choosing an appropriate family of hashes, Bob can send a hash key determining the hash function $h$, and for each $j$ send the corresponding hash $h(x_{1j},x_{2j},\ldots,x_{2^\sigma j})$ consisting of fewer bits.
If the hash function is sufficiently unpredictable, there is only a small risk that Alice fails to detect the corrupted channels. This, of course, has to be taken into consideration when bounding the reliability, but since the overall communication complexity of Protocol~\ref{prot:active} would still be exponential, we refrain from doing a thorough analysis.

\section{A polynomial time protocol}
Having now established that PRA cannot be solved in a single round, but can in three -- albeit in exponential time -- we propose a polynomial time protocol. While it requires four rounds rather than three, it does show the feasibility of the problem.
For the protocol, we need the following hash families defined by \cite{WegmanCarter,Stinson1994}.
\begin{defi}
  A family of functions $\mathcal{H}=\{h\colon X\rightarrow Y\}$ is called $\varepsilon$-strongly universal if it satisfies
  \begin{enumerate}
    \item for any $x\in X$ and any $y\in Y$, $\Pr_{h\in\mathcal{H}}[h(x)=y]=\frac{1}{|Y|}$
    \item for any $x_1,x_2\in X$ with $x_1\neq x_2$ and any $y_1,y_2\in Y$, $\Pr_{h\in\mathcal{H}}[h(x_1)=y_1, h(x_2)=y_2]\leq\frac{\varepsilon}{|Y|}$
  \end{enumerate}
  If $\varepsilon=\frac{1}{|Y|}$, then $\mathcal{H}$ is called strongly universal.
\end{defi}
Our proposed solution is given in Figure~\ref{prot:hash} on page~\pageref{prot:hash}.

\begin{prop}\label{prop:fourRound}
  After executing Protocol~\ref{prot:hash}, Alice and Bob will output the same key with probability at least $(1-\varepsilon)^{n-1}$.
\end{prop}
\begin{IEEEproof}
  In order for Alice and Bob to output two different keys, there must be some channel $i\in\mathcal{T}$ such that $\tilde{x}_i\neq x_i$. It is clear that such a channel is controlled by the adversary.
  Consider a single channel $i$ where $\tilde{x}_i\neq x_i$. The only way for $i$ to be added to $\mathcal{T}$ in step~\ref{item:checkHashes} is if the adversary given the knowledge that $h(x_i)=y_i$ has managed to find $(\tilde{x}_i,\tilde{y}_i)$ with $\tilde{x}_i\neq x_i$ such that $h(\tilde{x}_i)=\tilde{y}_i$. By the $\varepsilon$-strong universality of $\mathcal{H}$ this happens with probability
  \begin{align*}
    \Pr_{h\in\mathcal{H}}[h(\tilde{x}_i)=\tilde{y}_i&\mid h(x_i)=y_i]\\
    &=\frac{\Pr_{h\in\mathcal{H}}[h(\tilde{x}_i)=\tilde{y}_i, h(x_i)=y_i]}{\Pr_{h\in\mathcal{H}}[h(x_i)=y_i]}
    \leq \varepsilon
  \end{align*}
  regardless of the choice of $\tilde{y}_i$ and $\tilde{x}_i\neq x_i$.

  Thus, the probability that Alice and Bob discover that the $i$'th channel has been tampered with is at least $1-\varepsilon$. If the adversary tampers with $t^\ast\leq t$ channels, the probability that they are all discovered is at least $(1-\varepsilon)^{t^\ast}\geq (1-\varepsilon)^t=(1-\varepsilon)^{n-1}$ since the hash functions $h_i$ are sampled independently.
\end{IEEEproof}
\begin{figure*}[bt]
  \small
  \rule{\linewidth}{1.5pt}
  \begin{center}
    \unskip\textbf{Protocol~\ref{prot:hash}}
  \end{center}
  \unskip
  \begin{enumerate}
    \item Alice samples a random string $x\in\mathbb{F}_q^\ell$ uniformly at random, and creates an additive sharing $x=\sum_{i=1}^n x_i$. She then samples $h_i\in\mathcal{H}$ independently and uniformly to compute $y_i=h_i(x_i)$ for $i=1,2,\ldots,n$.

    She routes $(x_i,y_i)$ via the $i$'th channel.

    \item Bob receives the possibly altered messages $(\tilde{x}_{i},\tilde{y}_i)$, and sends a set $\mathcal{B}$ containing the blocked channels to Alice via the public channel.

    \item Alice transmits a description of each $h_i$ via the public channel.

    \item\label{item:checkHashes} Bob computes the set
    \begin{equation*}
      \mathcal{T}=\{i\in\{1,2,\ldots,n\}\mid h_i(\tilde{x}_i)=\tilde{y}_i\},
    \end{equation*}
    and sends this to Alice via the public channel. He outputs $k_B=\sum_{i\in\mathcal{T}} \tilde{x}_i$. When receiving $\mathcal{T}$, Alice computes and outputs $k_A=\sum_{i\in\mathcal{T}} x_i$.
  \end{enumerate}
  \rule{\linewidth}{1.5pt}
  \caption{Four-round key-exchange against active computationally bounded adversaries.
      After performing this PRA protocol for $n$ channels, Alice and Bob will share a key unknown to the adversary. Before executing the protocol, they have agreed on an $\varepsilon$-strongly universal family of hash functions $\mathcal{H}=\{h\colon\mathbb{F}_q^{\ell}\rightarrow Y\}$.
  }\label{prot:hash}
\end{figure*}
\begin{prop}\label{prop:protHashSecure}
  When using an $\varepsilon$-strongly universal family of hash functions, Protocol~\ref{prot:hash} is $\delta$-reliable for $\delta=1-(1-\varepsilon)^{n-1}$.
\end{prop}
\begin{IEEEproof}
  The reliability follows from Proposition~\ref{prop:fourRound}. Thus, we only need to prove that it is perfectly private.

  We use the same strategy as in proof of Proposition~\ref{prop:threeRoundSecure}. That is, we focus on the difference $E=K_B-K_A$, which is in this protocol determined by $\mathcal{T}$, $(X_i)_{i=1}^n$, and $(\tilde{X}_i)_{i=1}^n$. The set $\mathcal{T}$ is in turn determined by $(H_i)_{i\neq i^\ast}$ and $(\tilde{Y}_i)_{i\neq i^\ast}$, the latter being determined completely from $(Y_i)_{i\neq i^\ast}$ and $(X_i)_{i\neq i^\ast}$ when the input randomness of the adversary is fixed.
  Thus, conditioning on a value $E=e$ corresponds to a restricting to subset of tuples $\big( (X_i)_{i\neq i^\ast}, (H_i)_{i\neq i^\ast}\big)$. These determine the variables mentioned above uniquely.
  Like in the proof of Proposition~\ref{prop:threeRoundSecure}, each $X_i$ and each $H_i$ are sampled independently and uniformly at random, meaning that each tuple is equally likely to appear in the transcript.
  So we need only verify that such a transcript can correspond to any key $k_A$. Since $X_{i^\ast}$ is unknown to the adversary, this implies that $K_A$ can take any possible value and still agree with the transcript of the adversary as claimed.
\end{IEEEproof}
Before moving on to a specific implementation of Protocol~\ref{prot:hash}, we analyse its properties in general.
During Protocol~\ref{prot:hash}, the number of $q$-bits Alice sends in the first round is $n(\ell+\log_q|Y|)$ and in the third $n\log_q|\mathcal{H}|$. Bob sends the sets $\mathcal{B}$ and $\mathcal{T}$, each of size at most $n$, which means that both can be encoded in $n\log_q(2)$ $q$-bits.

Based on Proposition~\ref{prop:fourRound}, we can give a condition on the possible hash families if a specific maximal probability $\delta$ of protocol failure is desired. Namely, to get $\delta$-reliability, we must choose $\varepsilon$ such that
\begin{equation}\label{eq:epsBound}
  1-(1-\varepsilon)^{n-1}\leq \delta
  \quad\Longleftrightarrow\quad
  \varepsilon\leq 1-(1-\delta)^{\frac{1}{n-1}}.
\end{equation}

\subsection{A specific instantiation}\label{sec:hashInstance}
We now consider the parameters of Protocol~\ref{prot:hash} if the hash family employed is based on the multilinear hash family~\cite{CarterWegman1979,Lemire14}. Define for each $\mathbf{c}\in\mathbb{F}_q^{\ell+1}$ the map $h_{\mathbf{c}}\colon\mathbb{F}_q^\ell\rightarrow\mathbb{F}_q$ given by $h_{\mathbf{c}}(\mathbf{x})=c_1+\sum_{j=1}^\ell c_{j+1}x_j$. Then the multilinear family is $\mathcal{H}_{\mathsf{ML}}=\{h_{\mathbf{c}}\mid \mathbf{c}\in\mathbb{F}_q^{\ell+1}\}$, and this is a strongly universal family of hashes.

Now, observe that given an $\varepsilon$-strongly universal family $\mathcal{H}=\{h\colon X\rightarrow Y\}$, we can create another family
\begin{equation*}
  \mathcal{H}^\eta=\left\{h\colon X\rightarrow Y^\eta \;\middle|\;
  \begin{array}{l}
    h(x)=(h_1(x),h_2(x),\ldots,h_\eta(x))\\
    \text{and } h_i\in\mathcal{H},
  \end{array}
  \right\}
\end{equation*}
which is $\varepsilon^\eta$-strongly universal. This is the same idea as used in~\cite{Christensen19}.

If we employ $\mathcal{H}_{\mathsf{ML}}^\eta$, which is strongly universal (i.e. $\varepsilon=1/|\mathbb{F}_q^\eta|$), in Protocol~\ref{prot:hash}, we see from~\eqref{eq:epsBound} that
\begin{equation*}
  \frac{1}{|\mathbb{F}_q^\eta|}\leq 1-\sqrt[n-1]{1-\delta}
  \quad\Rightarrow\quad
  q\geq \left(\frac{1}{1-\sqrt[n-1]{1-\delta}}\right)^{\frac{1}{\eta}}
\end{equation*}
Thus, we have a lower bound on the required field size given $\varepsilon$, $n$, and $\eta$.
In order to get a simpler bound on the field size $q$, we use Bernoulli's inequality $(1-x)^r\leq 1-rx$ when $x\leq 1$ and $r\in[0,1]$, see for instance \cite[Sec.\,2.4]{mitrinovic}. Thus, setting $q\geq \left(\frac{n-1}{\delta}\right)^{\frac{1}{\eta}}$, we obtain
\begin{equation*}
  \varepsilon=\frac{1}{|\mathbb{F}_q^\eta|}
  \leq\frac{\delta}{n-1}
  =1-\left(1- \frac{\delta}{n-1}\right)
  \leq 1-(1-\delta)^{\frac{1}{n-1}},
\end{equation*}
and comparing this to~\eqref{eq:epsBound}, we see that this field size gives the desired success probability in Protocol~\ref{prot:hash} when $\mathcal{H}_{\mathsf{ML}}^\eta$ is used. Note that an element of $\mathcal{H}_{\mathsf{ML}}^\eta$ can be described in $\eta(\ell+1)$ $q$-bits. This means that the total number of $q$-bits transmitted by Alice and Bob collectively is
\begin{align*}
  n(\ell\!+\!\eta)+n\eta(\ell\!+\!1)+2n\log_q(2) &\leq n(\ell\!+\!\eta) +n\eta(\ell\!+\!1) +2n\\
                                         &= n(\eta+1)(\ell+2)
\end{align*}
since $q\geq 2$. The resulting key has length $\ell$, and the rate is thereby at most
$n(\eta+1)(\ell+1)/\ell$,
which reduces to $\mathcal{O}(n)$ under the assumption that $\eta$ is $\mathcal{O}(1)$.

\section{Conclusion and open problems}
In this work, we defined a new variation on the problem of private message transmission called private randomness agreement. We showed that any solution requires at least two transmission rounds, and that there exists an exponential time protocol requiring three rounds. We then gave a four-round protocol running in polynomial time.

The proposed solutions to PRA can be applied in QKD networks by considering the network as a graph containing $n$ vertex-disjoint paths. These paths correspond to the main channels in PRA, so Protocols~\ref{prot:active} and \ref{prot:hash} allow generation of a shared key with the security guarantees stated in Propositions~\ref{prop:threeRoundSecure} and \ref{prop:protHashSecure}. Compared to PMT, the PRA assumption tolerates a strictly more powerful adversary, allowing up to $t=n-1$ corrupted channels rather than $t<n/2$ in the PMT case. The cost is that a public, authenticated channel must be available.

An obvious question left open is whether the gap between the four rounds of Protocol~\ref{prot:hash} and the theoretical minimum of two rounds can be closed. Regardless of whether the answer is affirmative or negative, it will be an interesting contribution.

Another possible generalization is to explore the problem when allowing a small leakage, similar to $(\varepsilon,\delta)$-security in the PMT case. This would allow more flexibility in allowing a small loss of privacy, which might lead to different requirements on the minimal number of rounds. Such a generalization would perhaps not be reasonable to apply to the QKD problem, but other applications might have security requirements that are more relaxed.

Finally, the optimal transmission rate for PRA is unknown. For PMT it is shown in~\cite{Patra} that the transmission rate is lower bounded by $\Omega(n/(n-t))$ for two rounds or more, giving $\Omega(1)$ in the worst case.
It is unclear whether a constant rate is also possible for PRA, or if strictly more communication is needed in this case.

\bibliographystyle{ieeetr}
\bibliography{bibliography.bib}

\vfill

\end{document}